# Elastic, thermodynamic, electronic and optical properties of U$_2$Ti

M.Z. Hasan, M.M. Hossain, M.S. Islam, F. Parvin, A.K.M.A. Islam[*]

*Department of Physics, Rajshahi University, Rajshahi, Bangladesh*

**ABSTRACT**

An investigation of U$_2$Ti, a potentially safe and heavy metal-based storage material for radioactive tritium for fusion reactor, has been performed using pseudopotential density functional theory. The analysis of the elastic constants and other moduli calculated for the first time shows large anisotropy on elasticity and brittle behavior. A quasi-harmonic Debye model, which considers the vibrational contribution to the total free energy of the system, has been used to investigate the finite-temperature and finite-pressure thermodynamic properties of U$_2$Ti. The electronic band structure reveals metallic conductivity and the major contribution comes from U-5$f$ states. By analyzing the optical spectra, the origin of the various structures is also explained in terms of the calculated electronic structure. Further the reflectivity spectrum shows that the material is perfect reflector within the energy range 8-12.5 eV.

## 1. Introduction

Usually the most widely used and safest method of hydrogen isotope storage is solid state materials. The hydrogen storage behavior of heavy metal uranium has been studied to some extent [1-3]. But due to many disadvantages of using U, efforts have been made to improve hydrogen storage property by alloying [4-9]. As titanium exhibits high durability to powdering on hydrogenation, the U$_2$Ti intermetallic compound is likely to possess an excellent durability to powdering. Yamamoto *et al.* [10] have studied the hydrogen absorption-desorption behavior of U$_2$Ti, and its application as a storage material. The investigation of this behavior has been carried out over a temperature range of 298-973K, and at hydrogen pressure below 0.1 GPa. In view of all these it is apparent that the solid U$_2$Ti could serve as a potentially safe and heavy metal-based storage material for radioactive tritium for fusion reactor.

A fundamental understanding of the physico-chemical properties of U$_2$Ti is thus of interest. Chattaraj *et al.* [11] have recently carried out some works on structural and electronic properties of U, Ti, and U$_2$Ti. But they have not done any work on thermodynamic properties except deriving a value of the coefficient of electronic specific heat from the DOS at Fermi level using a theoretical expression. In their work these authors [11] verified that the systems are non-magnetic in nature.

To the best of our knowledge no theoretical study has yet been reported about thermodynamic, elastic and optical properties of U$_2$Ti. It is well-known that the elastic properties are essential for the understanding of the macroscopic mechanical properties of U$_2$Ti crystals because they are related to various fundamental solid state and thermodynamic properties. So, in the present work, we proceed with a description of the elastic, thermodynamic, electronic and optical properties of U$_2$Ti. Further the parameters of optical properties (absorption, conductivity, reflectivity, refractive index, energy-loss spectrum and dielectric function) will be calculated and discussed.

---

[*] Corresponding author. Tel.: +88 0721 750980; fax: +88 0721 750064.
 *E-mail address*: azi46@ru.ac.bd (A.K.M.A. Islam).

## 2. Computational Details

The first-principles *ab-initio* calculations are performed using the CASTEP code [12] in the framework of density functional theory (DFT) with generalized gradient approximation (GGA) and the Perdew-Burke-Ernzerhof (PBE) as exchange functional [13]. The interactions between ion and electron are represented by ultrasoft Vanderbilt-type pseudopotentials for U and Ti atoms [14]. An effect of spin-orbit coupling is expected to arise in $U_2Ti$ due to the heavy element U. But unfortunately we could not include such contribution in our calculations as the available CASTEP code does not allow any spin-orbit term in the DFT Hamiltonian and therefore we limit our calculations with such approximations.

The valence electron configurations of U and Ti were set to $6s^2 6p^6 6d^1 5f^3 7s^2$ and $3d^3 4s^1$, respectively. The elastic constants are calculated by the 'stress-strain' method. All the calculating properties for $U_2Ti$ used a plane-wave cut-off energy 900 eV and 9×9×13 Monkhorst-Pack [15] grid for the sampling of the Brillouin zone. Geometry optimization is conducted using convergence thresholds of $5\times10^{-6}$ eV atom$^{-1}$ for the total energy, 0.01 eVÅ$^{-1}$ for the maximum force, 0.02 GPa for maximum stress and $5\times10^{-4}$ Å for maximum displacement.

## 3. Results and discussion

### 3.1. Structural properties

$U_2Ti$ crystallizes in the *P6$_3$/mmm* space group and has 3 atoms in one unit cell. We have performed the geometry optimization as a function of the normal stress by minimizing the total energy of $U_2Ti$. The crystal structure of $U_2Ti$ is illustrated in Fig. 1. The optimized parameters for $U_2Ti$ are compared with available theoretical calculation [11] and experiment [16] (Table 1). The calculated results are in good agreement with experiment. It has been pointed out by Chattaraj *et al.* [11] that no significant difference in terms of lattice parameters is found if one includes spin-orbit calculation.

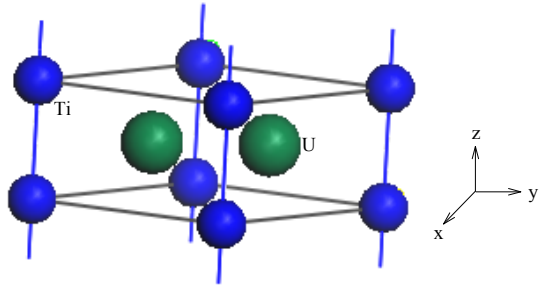

Fig. 1. Crystal structure of $U_2Ti$.

**Table 1.** The optimized structural parameters for $U_2Ti$.

| $T$ (K) | $a$ (Å) | $c$ (Å) | $c/a$ | $V_0$ (Å$^3$) | Ref |
|---|---|---|---|---|---|
| 0 | 4.782 | 2.822 | 0.59 | 55.87 | This |
| 0 | 4.773 | 2.815 | 0.59 | 55.54 | [11] |
| 298 | 4.828 | 2.847 | 0.59 | 57.47 | Expt.[16] |

*3.2. Mechanical properties*

To study the mechanical properties of $U_2Ti$ at $T = 0K$ and $P = 0$ GPa, the independent elastic constants $C_{ij}$, bulk modulus $B$, shear modulus $G$, Young's modulus $E$, and Poisson's ratio $\nu$ have been calculated. Theoretical details on elastic constants can be found elsewhere [17,18]. For hexagonal $U_2Ti$, five independent elastic constants are calculated (in GPa) as: $C_{11}= 285$, $C_{12}= 74$, $C_{13}= 28$, $C_{33}= 300$, $C_{44}= 129$. Unfortunately no other theoretical or experimental data exist to check our computed elastic constants.

The theoretical polycrystalline moduli for $U_2Ti$ may be computed from the set of independent elastic constants. Hill [19] proved that the upper and lower limits of the true polycrystalline constants are expressed from Voigt and Reuss equations. So the polycrystalline moduli are defined as the average values of the Voigt ($B_V$, $G_V$) and Reuss ($B_R$, $G_R$) moduli. According to Hill's observation, the value of bulk modulus (in GPa) $B = B_H = (B_V + B_R)/2$ (Hill's bulk modulus), where $B_V$ and $B_R$ are the Voigt's and the Reuss's bulk modulus respectively. The value of shear modulus $G = G_H = (G_V + G_R)/2$ (Hill's shear modulus), where $G_V$ and $G_R$ are the Voigt's and the Reuss's shear modulus respectively. The expressions for Voigt and Reuss moduli can be found in Ref. [20]. Using the two formulas: $E = 9BG/(3B + G)$ and $\nu = (3B - E)/6B$, the polycrystalline Young's modulus $E$ (in GPa) and the Poisson's ratio $\nu$ are then obtained. The values are: $B = 125$, $G = 121$, $E = 274$ (all in GPa); $\nu = 0.13$.

The elastic anisotropy of crystal, defined by the ratio $A = 2C_{44}/(C_{11} - C_{12})$ [21], yields a value of 1.22 for $A$. The factor $A = 1$ represents complete isotropy, while value smaller or greater than this measures the degree of anisotropy. Therefore, $U_2Ti$ shows anisotropic behavior. The parameter $k_c/k_a = (C_{11}+C_{12}-2C_{13})/(C_{33}-C_{13})$ expresses the ratio between linear compressibility coefficients of hexagonal crystals [21]. From our data the value of $k_c/k_a$ (= 1.11) indicates that the compressibility for $U_2Ti$ along $c$ axis is greater than along $a$ axis. According to Pugh's criteria [22], a material behaves in a ductile manner, if $G/B < 0.5$, otherwise it should be brittle. A value of 0.96 for $U_2Ti$ thus indicates its brittle behavior.

*3.3. Thermodynamic properties*

We investigated the thermodynamic properties of $U_2Ti$ by using the quasi-harmonic Debye model, the detailed description of which can be found in literature [23]. For this we first derive *E-V* data obtained from Birch-Murnaghan equation of state [24] using zero temperature and zero pressure equilibrium values, $E_0$, $V_0$, $B_0$, based on DFT method. Then in order to get different thermodynamic properties at finite-temperature and finite-pressure, we apply the quasi-harmonic Debye model, in which the non-equilibrium Gibbs function $G^*(V; P, T)$ can be written in the form [23]:

$$G^*(V;P,T) = E(V) + PV + A_{vib}[\Theta(V);T] \qquad (1)$$

where $E(V)$ is the total energy per unit cell, $PV$ corresponds to the constant hydrostatic pressure condition, $\Theta(V)$ is the Debye temperature, and $A_{vib}$ is the vibrational term, which can be written using the Debye model of the phonon density of states as [23]:

$$A_{vib}(\Theta,T) = nkT\left[\frac{9\Theta}{8T} + 3\ln(1-\exp(-\Theta/T)) - D\left(\frac{\Theta}{T}\right)\right] \qquad (2)$$

where $n$ is the number of atoms per formula unit, $D(\Theta/T)$ represents the Debye integral.

The non-equilibrium Gibbs function $G^*(V; P, T)$ can be minimized with respect to volume $V$ to obtain the thermal equation of state $V(P, T)$ and the chemical potential $G(P, T)$ of the corresponding phase. Other macroscopic properties can also be derived as a function of $P$ and $T$ from standard thermodynamic relations [23]. Here we computed the normalized volume, bulk modulus, specific heats, Debye temperature and volume thermal expansion coefficient (VTEC) at different temperatures and pressures.

Fig. 2 shows the pressure dependent normalized volume $V/V_0$ and bulk modulus $B$ for $U_2Ti$ at different temperatures. It is seen that the unit cell volume decreases smoothly with increasing pressure which indicates that the crystal structure of U$_2$Ti is stable up to 60 GPa. Inset shows the temperature variation of bulk modulus for three different pressures. We see that $B$ decreases slightly with increasing temperature. As an example at zero pressure, $B$ drops by ~7.5% in a slightly nonlinear manner from 0 to 1200K.

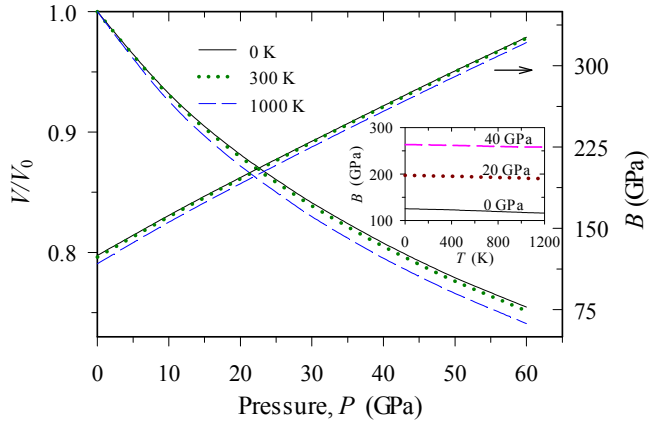

**Fig. 2.** The pressure variation of $V/V_0$ and $B$ at different temperatures. *Inset*: Temperature dependence of $B$ at three different pressures.

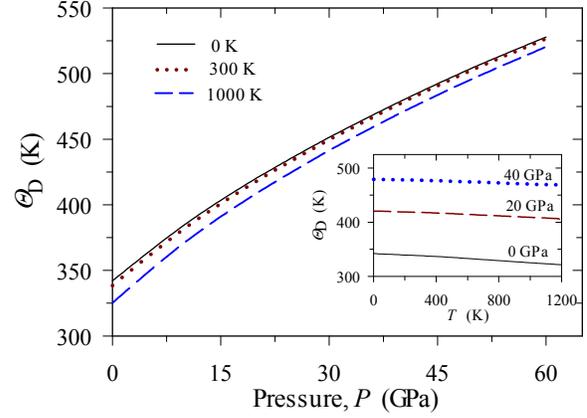

**Fig. 3.** The pressure dependence of $\Theta_D$ at different temperatures. *Inset*: Temperature variation of $\Theta_D$ at different pressures.

Fig. 3 shows the pressure dependence of Debye temperature $\Theta_D$ at three different temperatures. Inset shows the temperature variation of $\Theta_D$ at three different pressures. We can see that $\Theta_D$ increases with increase in pressure, while it decreases smoothly with increasing temperature. We know that $\Theta_D$ is related to the maximum thermal vibration frequency of a solid. Due to this relationship, the variation of $\Theta_D$ with temperature and pressure also reveals the changeable vibration frequency of the particles in U$_2$Ti.

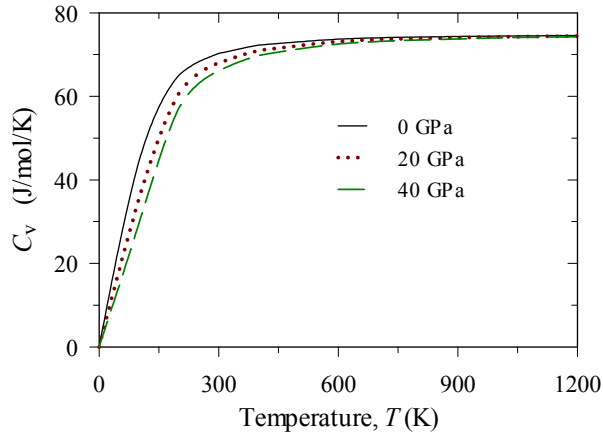

**Fig. 4.** The temperature dependence of specific heat capacity $C_V$ at different pressures.

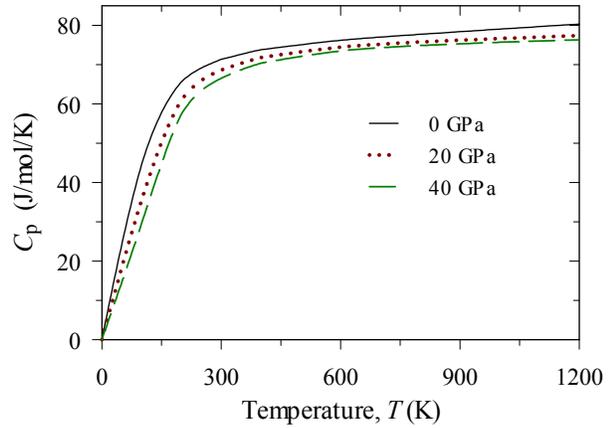

**Fig. 5.** The temperature dependence of specific heat capacity $C_P$ at different pressures.

The temperature dependence of constant-volume specific heat $C_V$ and constant-pressure specific heat $C_P$ of U$_2$Ti are shown in Fig. 4 and Fig. 5, respectively. In fact, the heat capacities increase with increasing temperature, because phonon thermal softening occurs when the temperature increases. The difference between $C_P$ and $C_V$ for $U_2Ti$ is very small and is given by $C_P - C_V = \alpha_V^2(T) BTV$, where $\alpha_V$ is the volume thermal expansion coefficient. The difference is due to the thermal expansion caused by anharmonicity effects. In the low-temperature limit, the specific heat exhibits the Debye $T^3$ power-law behavior and at high temperature, both approach the classical asymptotic limit of $C_V = 3nNk_B = 73.8$ J/mol/K. The constant pressure heat capacity data at low temperature indicate monotonically increasing temperature dependence [25]. The electronic and phonon contribution to the specific heat is usually obtained by fitting the measured specific heat at low temperatures to $C_P/T = \gamma + \beta T^2$, where the parameters $\gamma$ and $\beta$ are the electronic and phonon contributions, respectively. In the absence of measured specific heat, the coefficient $\gamma$ may be calculated from the DOS, $N(E_F)$ at Fermi level using $\gamma = \pi^2 k_B^2 N(E_F)/3$. Using our $N(E_F)$ value (see Fig.7b), we obtain $\gamma = 0.0111$ J/mol/K$^2$ compared to 0.01475 J/mol/K$^2$ obtained by Chattaraj *et al.* [11].

The temperature dependence of the volume thermal expansion coefficient, $\alpha_V$ [31] is calculated as a function of both temperature and pressure (Fig. 6). At low temperature region ($T < 300$K), the coefficient increases rapidly with increasing temperature and then increases smoothly and slowly at high temperatures. It is well-known that the thermal expansion coefficient is inversely related to the bulk modulus of a material. Since there is no measured value of VTEC ($\alpha_V$) for $U_2Ti$, we seek to check the reliability of our calculation through $\alpha$-U for which measured and theoretical values are available. Our calculation provides VTEC as $3.76 \times 10^{-5}$ K$^{-1}$ at 300K. Assuming, linear thermal expansion coefficient = $\alpha_V/3$, our result of $12.5 \times 10^{-6}$ K$^{-1}$ at 300K is in very good agreement with the calculated (spin orbit) value of $12.8 \times 10^{-6}$ K$^{-1}$ [26] and measured value $12.6 \times 10^{-6}$ K$^{-1}$ [27] or $14.3 \times 10^{-6}$ K$^{-1}$ [28]. This would put some confidence in the methods we are using for $U_2Ti$.

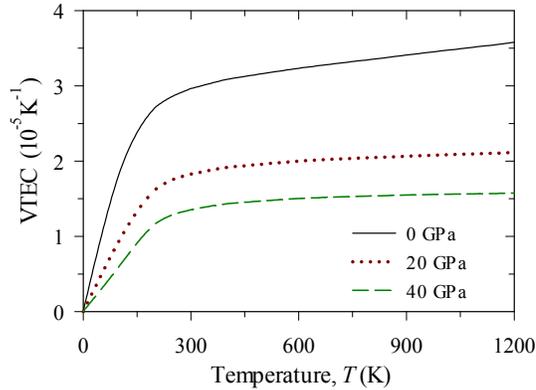

**Fig. 6.** The temperature dependence of VTEC $\alpha_V$ at different pressures.

*3.4. Electronic properties*

The energy band structure and density of states (DOS) of $U_2Ti$ are shown in Figs. 7 despite limitation due to neglect of spin-orbit (SO) correction in our calculation as mentioned earlier. Chattaraj *et al*. [11] discussed their results on U metal and alloy. For both U and $U_2Ti$, the nature of the DOS spectrum is found similar below the Fermi level. However, a substantial difference in the DOS spectrum is observed using SO and NSO approach above the Fermi level. For both $\alpha$-U and $U_2Ti$, the inclusion of spin-orbit effect results in narrowing the bandwidth and this effect is more prominent above the Fermi level. Unlike this, the DOS of Ti does not show any significant effect due to SO incorporation [11]. Our calculation shows that the band structure of $U_2Ti$ is of metallic conductivity, because there are many bands crossing the Fermi level. These energy bands are mainly from the U 5$f$ states around the Fermi energy. The dominating contribution on the conductivity originally comes from the U 5$f$ states. No bands are observed below the Fermi level within the energy range -4.82 to -15.42 eV. The lowest energy bands below the Fermi level (-15 to -22eV, not shown here) are mainly dominated from the U 6*p* states. Below the energy range (0 to -5eV), the dominating contribution arises from the U 6*d*/6*p* and Ti 3*d*/4*s*.

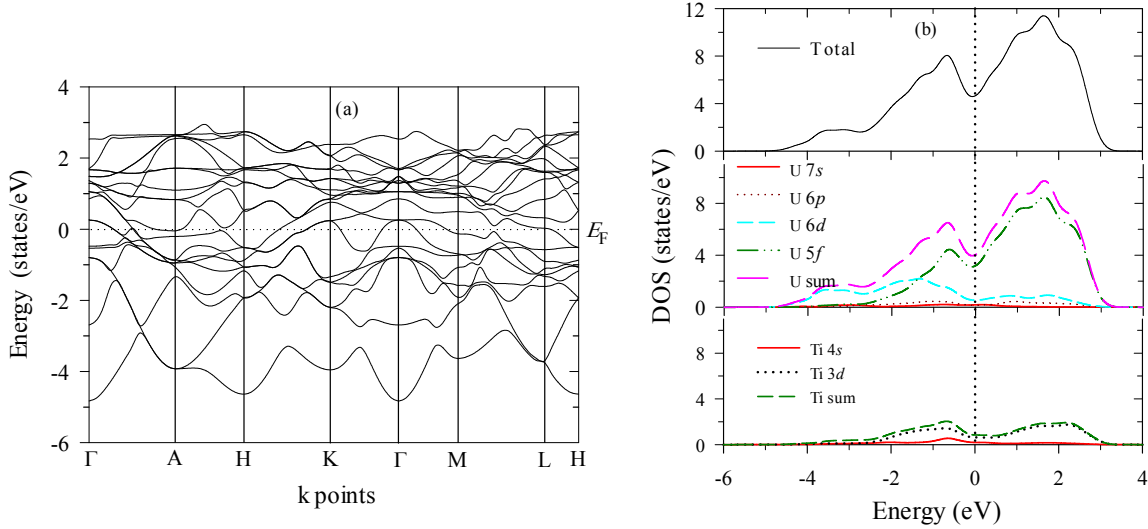

**Fig. 7.** (a) Band structure of U$_2$Ti, and (b) total and partial DOSs for U$_2$Ti.

*3.5. Optical properties*

The study of the optical functions helps to give a better understanding of the electronic structure. These may be obtained from the complex dielectric function, $\varepsilon(\omega) = \varepsilon_1(\omega) + i\varepsilon_2(\omega)$. The imaginary part $\varepsilon_2(\omega)$ is obtained from the momentum matrix elements between the occupied and the unoccupied electronic states and calculated directly using [29]:

$$\varepsilon_2(\omega) = \frac{2e^2\pi}{\Omega\varepsilon_0} \sum_{k,v,c} \left|\psi_k^c |u.r| \psi_k^v\right|^2 \delta\left(E_k^c - E_k^v - E\right) \quad (1)$$

where $u$ is the vector defining the polarization of the incident electric field, $\omega$ is the light frequency, $e$ is the electronic charge and $\psi_k^c$ and $\psi_k^v$ are the conduction and valence band wave functions at $k$, respectively. The real part is derived from the imaginary part $\varepsilon_2(\omega)$ by the Kramers-Kronig transform. All other optical constants, such as refractive index, absorption spectrum, loss-function, reflectivity and conductivity (real part) are those given by Eqs. 49 to 54 in ref. [29].

Fig. 8 shows the optical functions of U$_2$Ti calculated for photon energies up to 30 eV for polarization vectors [100] and [001]. We have used a 0.5 eV Gaussian smearing for all calculations. This smears out the Fermi level, so that k-points will be more effective on the Fermi surface.

The U$_2$Ti has an absorption band, similar for both the polarization vectors, in the low energy range due to its metallic nature (Fig. 8(a)). The spectrum from [100] polarization rises rapidly and then decreases slowly till ~9.2 eV and then sharply falls to zero. There is no absorption band within the energy range 9.2-14.7 eV and it has two peaks at about 17.4 and 20.7 eV. The origin of peaks is due to the interband transition from the Ti *d* to the U *f* states. The spectrum due to polarization [001] shows a sharper peak and then has a shoulder on the right.

Since the material has no band gap as evident from band structure, the photoconductivity starts with zero photon energy for both polarization vectors as shown in Fig. 8 (b). Moreover, the photoconductivity and hence electrical conductivity of a material increases as a result of absorbing photons [30]. Fig. 8 (c) shows the reflectivity spectra of U$_2$Ti (roughly similar for both the polarizations) as a function of photon energy. The spectrum shows that the material is a perfect reflector within the energy range 8-12.5 eV. This kind of nonselective characteristic in the wide energy range implies that $U_2Ti$ would be a better candidate material as a coating to avoid solar heating.

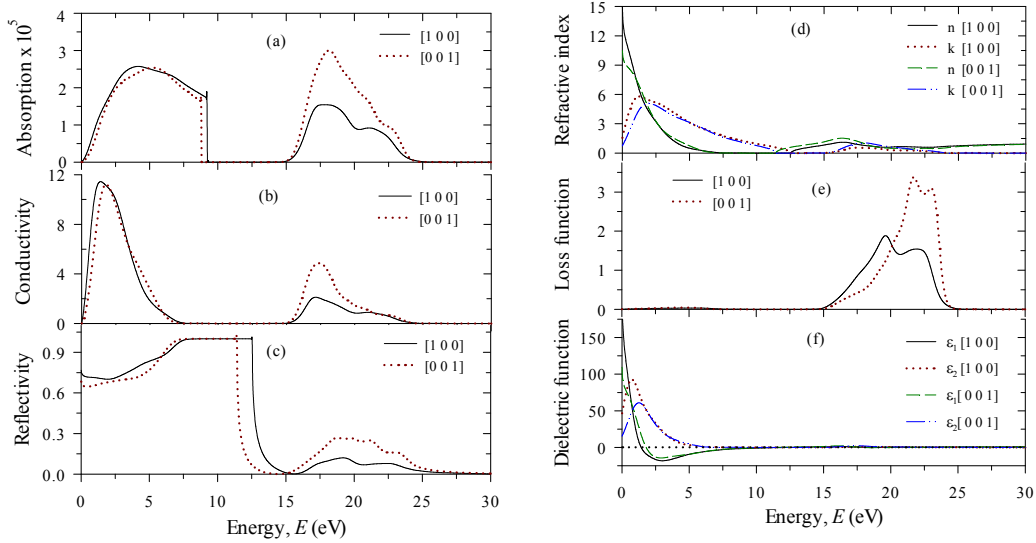

**Fig. 8.** The optical functions (a) absorption, (b) conductivity, (c) reflectivity, (d) refractive index, (e) loss function, and (f) dielectric function of $U_2Ti$ for two polarization vectors.

The refractive index and extinction coefficient are illustrated in Fig. 8 (d). The static refractive index of $U_2Ti$ is found to have the value ~13.5. In Fig. 8 (e), the electron energy loss function describing the energy loss of a fast electron traversing a material is usually large at the plasma frequency [31]. Prominent peak is found at 19.6 eV, which indicates rapid reduction in the reflectance. The peak which is much enhanced for [001] polarization vector is shifted towards right. The imaginary and real parts of the dielectric function are displayed in Fig. 8 (f). It is observed that the real part $\varepsilon_1$ vanishes at about 8 eV. $U_2Ti$ exhibits metallic reflectance characteristics in the range of $\varepsilon_1 < 0$. The peak of the imaginary part of the dielectric function is related to the electron excitation. For the imaginary part of $\varepsilon_2$, the peak for < 1 eV is due to the intraband transitions.

## 4. Conclusion

First-principles calculations based on density functional have been used to study the structural, elastic, thermodynamic, electronic and optical properties of $U_2Ti$. The analysis shows large anisotropy on elasticity and a brittle behavior. The finite-temperature and finite-pressure thermodynamic properties of $U_2Ti$ are evaluated through the use of quasi-harmonic Debye model, which considers the vibrational contribution to the total free energy of the system. Bulk modulus is seen to increase rapidly with pressure but decreases slightly with increasing temperature. The Debye temperature $\Theta_D$ is found to increase with increase in pressure, while it decreases smoothly with increasing temperature. Since $\Theta_D$ is related to the maximum thermal vibration frequency, the variation of $\Theta_D$ with temperature and pressure reveals the changeable vibration frequency of the particles in $U_2Ti$. The heat capacities increase with increasing temperature, which shows that phonon thermal softening occurs when the temperature increases. An estimation of the electronic contribution to specific heat has also been made.

The band structure of $U_2Ti$ shows its metallic conductivity and the major contribution originates from U-5$f$ states. Further, the optical properties, e.g. absorption, conductivity, reflectivity, refractive index, energy-loss spectrum and dielectric function of $U_2Ti$ from two polarization vectors are analyzed. The reflectivity spectrum shows that the material would be an excellent reflector within the energy range 8-12.5 eV.